\documentclass[12pt]{article}
\usepackage{amssymb}
\usepackage{graphicx,color}

\def\appendix{{\newpage\section*{Appendix}}\let\appendix\section%
        {\setcounter{section}{0}
        \gdef\thesection{\Alph{section}}}\section}

\newcommand{\be}{\begin{equation}}
\newcommand{\ee}{\end{equation}}
\newcommand{\bear}{\begin{eqnarray}}
\newcommand{\eear}{\end{eqnarray}}
\newcommand{\ba}{\begin{array}}
\newcommand{\ea}{\end{array}}

\begin{document}

\vspace{4mm}

\begin{center}
{
{\Large \bf  Hawking Temperature in Taub-NUT (A)dS spaces\\
via the Generalized Uncertainty Principle}\vspace{15mm}

{\large Seyen Kouwn$^{a,}$\footnote{E-mail: seyen@skku.edu},
Chong Oh Lee$^{b,}$\footnote{E-mail: cohlee@sciborg.uwaterloo.ca}, and
Phillial Oh$^{a,}$\footnote{E-mail: ploh@skku.edu}}

\vspace{10mm} {\em $^a$  Department of Physics and Institute of Basic Science,\\
Sungkyunkwan University, Suwon 440-746 Korea \\
 $^b$ Department of Physics and Astronomy, University of Waterloo,\\
Waterloo, Ontario, N2L 3G1, Canada}

}
\end{center}

\vspace{7mm}

\begin{abstract}
Using the extended forms of the Heisenberg uncertainty principle
from string theory and the non-commutative geometries, we derive the
Hawking temperature of the Taub-Nut-(A)dS black hole. We show that
the generalized versions of the Heisenberg uncertainty principle
increase the Hawking temperature as in the case of the
Schwarzschild-(A)dS black hole.
\end{abstract}

\newpage
\setcounter{equation}{0}
\section{Introduction}

The Taub-NUT metric was
suggested in
Ref.~\cite{Taub:1950ez,Newman:1963yy}
in search for metrics with high
symmetry, and as a natural
generalization of the
Schwarzschild metric. It was
also investigated in
Ref.~\cite{Misner:1963fr}
through  various analysis of
the metric. The generalized
version of the Taub-NUT-AdS
metric in arbitrary even
dimension was found in
Ref.~\cite{Awad:2000gg}. The
metric with NUT charge has
several special properties. For
example, the solutions of the
metric are not asymptotically
flat (AF) but asymptotically
locally flat
(ALF)~\cite{Hawking:1998jf,
Hawking:1998ct}. In contrast to
a common black hole, the
entropy is not just a  quarter
area at the
horizon~\cite{Hawking:1998jf,Hawking:1998ct}.
However, in spite of these
distinctive natures, such a
metric well satisfies the
AdS-CFT
correspondence~\cite{Mann:1999pc,Clarkson:2002uj,Lee:2008uh}.

The basic ideas of the
generalized uncertainty
principle (GUP) started from
considering how to construct a
gravity theory at the quantum
limit~\cite{Wigner:1957ep,Salecker:1957be}.
The GUP appeared in several
contexts. In the context of
string theory, the GUP appeared
in
~\cite{Veneziano:1986zf,Gross:1987ar,Amati:1988tn}
and in particular it was
explicitly showed that the
Heisenberg uncertainty
principle (HUP) needs some
modifications due to a minimum
physical
length~\cite{Konishi:1989wk}.
It revived many investigations
in quantum gravity along these
directions
~\cite{Maggiore:1993rv,Ng:1993jb,AmelinoCamelia:1994vs,Garay:1994en,Chang:2001bm}.
It was found that the space
noncommutativity is as
essentially similar concept as
the GUP
~\cite{Kempf:1994su,Nozari:2006bi},
the gravitational interaction
of the photon and the particle
can lead to the
GUP~\cite{Adler:1999bu}, and
the GUP increases the Hawking
temperature and decreases the
entropy~\cite{Cavaglia:2004jw}.
The GUP  also provides a useful
scheme to study the
thermodynamic quantities of
black hole. One can resolve the
quantum corrections to black
hole entropy without much
difficulty~\cite{Setare:2004sr,Medved:2004yu,Setare:2005sj,Ko:2006mr},
test the stability of the black
hole~\cite{Custodio:2003jp,Kim:2007hf},
and investigate thermal
properties in quantum
gravitational
effects~\cite{Adler:2001vs,Bolen:2004sq,Myung:2006qr,Park:2007az}.

It was discussed in~\cite{Bolen:2004sq} that only
Schwarzschild-(A)dS-like black holes seem to allow a holographic
dual theory since the generalized forms of the HUP in the high
energy regime lead to the thermodynamic quantities of
Schwarzschild-(A)dS (S-(A)dS). Since there is an equivalence between
a gravitational theory in the bulk and a conformal field theory on
the boundary in the Taub-Nut-(A)dS
spaces~\cite{Mann:1999pc,Clarkson:2002uj,Chang:2001bm},  one
intriguing question is whether the thermodynamics of the
Taub-Nut-(A)dS metric can also be obtained from the extended forms
of the HUP. Another is whether  the extended versions of the HUP
increase its Hawking temperature as in the case of S-(A)dS-like
black hole \cite{Adler:2001vs,Cavaglia:2004jw,Park:2007az}.

In this paper, we address these questions adopting the method of
Ref. \cite{Bolen:2004sq}\footnote{ The  approach  of Ref.
\cite{Bolen:2004sq} was criticized on  a couple of points. For
example, In Ref. \cite{Scardigli:2006eb}, the Hawking temperature
for a large class of black holes including Schwarzschild-(A)dS black
hole was computed using only HUP and classical physics. In Ref.
\cite{Bambi:2007ty}, it was pointed out that a minimum in the
momentum uncertainty is absent due to the negative coefficient in
the dual form of the generalized uncertainty principle (see the
coefficient $\beta^2$ in Eq. (\ref{EUP}) and (\ref{dbeta})) in the
case of Schwarzschild-dS black hole. In the same paper, it was shown
that a positive value can be achieved  using only quantum mechanics
and general relativity through a simple gedanken experiments. But
these criticisms are based on heuristic arguments, and   do not
definitely exclude the present analysis. }. We show that the
thermodynamic of Taub-Nut-(A)dS metric can be explained with the
generalized Heisenberg uncertainty principle and the Hawking
temperature increases as in the case of the S-(A)dS black hole.

\setcounter{equation}{0}
\section{Hawking Temperature of Taub-NUT and Taub-NUT-AdS in 3+1 dimension}
In this section, we consider a
four-dimensional Taub-NUT with
or without a negative
cosmological constant and show
that the Hawking temperature of
the Taub-NUT black hole can be
obtained from the conventional
HUP but that of the
Taub-NUT-(A)dS black hole can
not be obtained from HUP.
\subsection{Four-dimensional Taub-NUT}\label{4DTaub-NUT}
The metric of the Taub-NUT black hole in 3+1 dimension is given
by~\cite{Newman:1963yy,Misner:1963fr}
\begin{eqnarray}\label{TNmetric}
ds^{2}=-f(r)\biggr[ dt + 2n \cos \theta d\phi \biggr]^{2}
       +\frac{dr^{2}}{f(r)} + (r^{2}+n^{2})d\Omega_{2}^{2} \,,
\end{eqnarray}
where $f(r)$ and $d\Omega_{2}^{2}$ are
\begin{eqnarray}\label{TNfr}
f(r)=\frac{r^{2}-n^{2}}{r^{2}+n^{2}}-\frac{2mr}{r^{2}+n^{2}}\,,
\hspace{1cm} d\Omega_{2}^{2} = d\theta^{2} + \sin^{2}\theta
d\phi^{2}\,.
\end{eqnarray}
Here, $n$ denotes the NUT
charge and $m$ is a geometric
mass. The Hawking temperature
$T_H$ arises from the imposed
condition which ensures
regularity in the Euclidean
time $t_{E}$ and radial
coordinate $r$
\bear\label{ghtemp}
\frac{1}{T_H}=\left.\frac{4\pi}{\partial_r
f(r)}\right|_{r=r_{+}}, \eear
where $1/T_H$ is the period of
Euclidean time $t_{E}$ and
$r_+$ is the radius of the
event horizon.

Then, the Hawking temperature
$T_H$ becomes
\begin{eqnarray}\label{TNT}
T_{H} = \frac{\hbar c}{4\pi r_+}\,.
\end{eqnarray}
The HUP and thermal properties of black hole also lead to the
Hawking temperature (\ref{TNT})~\cite{Adler:2001vs,Bolen:2004sq}. By
modeling a black hole as a black box with linear size $r_+$, the
uncertainty in the energy of the emitted quanta by the Hawking
effect is
\begin{eqnarray}\label{TNT1}
\Delta E \simeq c \Delta p  \simeq \frac{\hbar c}{r_{+}}\simeq
\frac{\hbar c}{\Delta x}\,.
\end{eqnarray}
$\Delta E$ can be identified as
the characteristic temperature
of the Hawking radiation.
Employing an appropriate
proportional constant $1/4\pi$,
we can reproduce the Hawking
temperature $T_{H}$
(\ref{TNT}). This implies that
the HUP in even the Taub-NUT
space well suffices to derive
the black hole temperature.
\subsection{Taub-NUT-AdS$_{4}$}\label{sub22}
The Taub-NUT-AdS black hole line element of 3+1 dimension is given by
\begin{eqnarray}\label{TNAmetric}
ds^{2}=-g(r)\biggr( dt + 2n \cos \theta d\phi \biggr)^{2}
       +\frac{dr^{2}}{g(r)} + (r^{2}+n^{2})d\Omega_{2}^{2} \,,
\end{eqnarray}
where $g(r)$ is \bear\label{TNAfr} g(r)=\frac{l^2 r^2-n^2
l^2+r^4+6n^2r^2-3n^4}{l^2 (r^2+n^2)}-\frac{2m r}{r^2+n^2}, \eear and
a negative cosmological constant $\Lambda$ is $\Lambda=-3/l^2$.
Using the same method as in the previous subsection, the
Hawking temperature with the black hole horizon $r_{+}$ is given by
\begin{eqnarray}\label{TNAT}
T_{H{\rm(AdS)}} = \frac{1}{4\pi} \biggr[ \frac{1}{r_{+}} +
\frac{3(n^2+r_{+}^2)}{l^2\,r_{+}} \biggr]\hbar c\,,
\end{eqnarray}
which has the feature that two thermodynamical limits may be
realized.  The direct evaluation of the above Hawking temperature
(\ref{TNAT}) is a difficult problem and beyond our present
abilities. But the  result of (\ref{TNAT}) was also obtained in Ref.
\cite{Astefanesei:2004kn} before and it is consistent with the
Hawking temperature obtained by the tunneling method of Ref.
\cite{Kerner:2006vu}.

In the limit of the Taub-NUT, $r_+^2\ll {l^2}/{3} + n^2$, since the
radius of the event horizon is much smaller in comparison to the
curvature radius of the AdS space, the temperature of the
Taub-NUT-AdS$_4$ black hole becomes, \bear T_{H{\rm(AdS)}}\simeq
\frac{\gamma\,\hbar c}{4\pi r_{+}},~~{\rm for} ~r_+^2\ll
\frac{l^2}{3} + n^2 \eear where $\gamma=1+3n^2/l^2$, which is a
dimensionless constant. Comparing this with the previous result
(\ref{TNT1}), we have \begin{eqnarray} \Delta E \simeq
\frac{\gamma\hbar c}{r_{+}} \,,\label{vacuum1}
\end{eqnarray}
where $\Delta E$ is again the characteristic energy associated with
the Hawking  temperature (2.9). From this, we infer HUP for
Taub-NUT-AdS$_4$ black hole can be obtained $\Delta x \Delta p \simeq
\gamma \hbar$. Note that for zero NUT charge, $n=0$, $\gamma=1$, and
HUP takes the standard form. The shifting of $\gamma$ for non-zero
$n$ is purely characteristic of Taub-NUT-AdS$_4$ solution of the
vacuum Einstein equation.

 In the AdS limit, $r_+^2 \gg
{l^2}/{3} + n^2$, the radius of
black hole event horizon
dramatically grows up in
comparison to the radius of
curvature of the AdS space so
that the temperature of the
(cosmological) horizon is
obtained by
\begin{eqnarray}\label{TNATlimAds}
T_{H{\rm(AdS)}}\simeq \frac{3
\hbar c}{4\pi
l^2}\,r_{+}\,,~~{\rm for}~r_+^2
\gg \frac{l^2}{3} + n^2
\end{eqnarray}
which cannot be explicitly
derived from the HUP. However,
the Hawking temperature
(\ref{TNAT}) of the
Taub-NUT-AdS$_4$ solution can
be obtained by replacing the
usual Heisenberg relation with
its extended versions, like
S-(A)dS-like black
hole~\cite{Bolen:2004sq}. The
details are discussed in
following sections.
\setcounter{equation}{0}
\section{Generalized Uncertainty Principle}
In this section we first
briefly mention the various
extended versions of HUP before
they are applied to the
derivation of the Hawking
temperature of the
Taub-NUT-AdS$_4$ space.

The GUP of the HUP is usually given
by~\cite{Adler:2001vs,Cavaglia:2004jw}
\begin{eqnarray}\label{GUP}
\Delta x \Delta p \ge \hbar \biggr[ 1 + \alpha^2
\l_{p}^2\frac{\Delta p^2}{\hbar^2} \biggr]\,,
\end{eqnarray}
where $l_{p}=(\hbar G/c^3)^{1/2}$ is the Plank length and $\alpha$ is
a dimensionless real constant of order one. For non-zero Taub-NUT
charge in (\ref{vacuum1}), we propose GUP in Taub-NUT AdS space as
\begin{eqnarray}\label{mGUP}
\Delta x \Delta p \ge \hbar \biggr[\gamma + \alpha^2
\l_{p}^2\frac{\Delta p^2}{\hbar^2} \biggr]\,,
\end{eqnarray}
The second
term in (\ref{mGUP}) gives rise to an absolute minimum in the
position uncertainty
\begin{eqnarray}\label{GUPxmin}
\Delta x \ge 2 \alpha l_{p}\sqrt{\gamma}\,,
\end{eqnarray}
and the uncertainty in the momentum is given by
\begin{eqnarray}\label{ineqGUPp}
\frac{\hbar\Delta x}{2\alpha^2 \l_{p}^2} \Biggr[ 1 -
\sqrt{1-\frac{4\gamma\alpha^2 \l_{p}^2}{(\Delta x)^2} } \Biggr] \leq
\Delta p  \leq \frac{\hbar\Delta x}{2\alpha^2 \l_{p}^2}
\Biggr[ 1 + \sqrt{1-\frac{4\gamma\alpha^2 \l_{p}^2}{(\Delta x)^2} } \Biggr]\,.\nonumber\\
\end{eqnarray}
Note that the GUP (\ref{GUP})
goes back to the usual HUP as
$2\alpha l_{p}\rightarrow 0$,
i.e. when $\Delta x \gg l_p$ or
$\alpha \rightarrow 0$. When
one considers the dual form of
the GUP, one gets another
generalized version of the
HUP~\cite{Kempf:1994su,Bolen:2004sq},
\begin{eqnarray}\label{EUP}
\Delta x \Delta p \ge \hbar \biggr[ \gamma + \beta^2 \frac{\Delta
x^2}{l_{p}^2} \biggr] \,,
\end{eqnarray}
where $\beta$ is a constant parameter. Then there exists an absolute
minimum in the momentum uncertainty
\begin{eqnarray}\label{EUPP}
\Delta p \ge \frac{2\hbar \beta\sqrt{\gamma}}{l_p},
\end{eqnarray}
and the uncertainty in the position is
\bear
\frac{l_p^2 \Delta p}{2 \hbar \beta^2} \Biggr[ 1 -
\sqrt{1-\frac{4\gamma\beta^2 \hbar^2}{l_p^2(\Delta p)^2}} \Biggr] \leq \Delta x
\leq \frac{l_p^2 \Delta p}{2 \hbar \beta^2} \Biggr[ 1 +
\sqrt{1-\frac{4\gamma\beta^2 \hbar^2}{l_p^2(\Delta p)^2}} \Biggr]\,.
\eear

By combining (\ref{GUP}) and
(\ref{EUP}), the symmetric
generalized uncertainty
principle is obtained as
\begin{eqnarray}\label{GEUP}
\Delta x \Delta p \ge \hbar \biggr[ \gamma + \alpha^2
\l_{p}^2\frac{\Delta p^2}{\hbar^2} + \beta^2 \frac{\Delta
x^2}{l_{p}^2} \biggr] \,.
\end{eqnarray}
 Inverting (\ref{GEUP}), the inequalities are given by
 \begin{eqnarray}\label{ineqx}
\frac{l_p^2 \Delta p}{2 \hbar \beta^2} \Biggr[ 1 -
\sqrt{1-\frac{4\beta^2 \hbar^2}{l_p^2(\Delta p)^2} \biggr[\gamma+\alpha^2
\l_{p}^2\frac{\Delta p^2}{\hbar^2}\biggr] } \Biggr] \leq \Delta x
\leq \frac{l_p^2 \Delta p}{2 \hbar \beta^2} \Biggr[ 1 +
\sqrt{1-\frac{4\beta^2 \hbar^2}{l_p^2(\Delta p)^2}
\biggr[\gamma+\alpha^2 \l_{p}^2\frac{\Delta p^2}{\hbar^2}\biggr] } \Biggr]\,,\nonumber\\
\end{eqnarray}
and
\begin{eqnarray}\label{ineqp}
\frac{\hbar\Delta x}{2\alpha^2 \l_{p}^2} \Biggr[ 1 -
\sqrt{1-\frac{4\alpha^2 \l_{p}^2}{(\Delta x)^2} \biggr[\gamma+\beta^2
\frac{(\Delta x)^2}{l_{p}^2}\biggr] } \Biggr] \leq \Delta p  \leq
\frac{\hbar\Delta x}{2\alpha^2 \l_{p}^2} \Biggr[ 1 +
\sqrt{1-\frac{4\alpha^2 \l_{p}^2}{(\Delta x)^2}
\biggr[\gamma+\beta^2 \frac{(\Delta x)^2}{l_{p}^2}\biggr] } \Biggr]\,,\nonumber\\
\end{eqnarray}
which lead to the absolute
minimum values of $\Delta x$
and $\Delta p $, respectively
by
\begin{eqnarray}\label{minx}
(\Delta x)^2 \ge \frac{4 \gamma\alpha^2 l_p^2}{1-4 \alpha^2
\beta^2}\,,\hspace{1cm} (\Delta p)^2 \ge \frac{4
\hbar^2\gamma\beta^2/l_p^2}{1-4 \alpha^2 \beta^2}\,.
\end{eqnarray}
The inequalities~(\ref{ineqx}) and~(\ref{ineqp}) are required to be
real so that an additional condition is given by
\bear
\beta^2<\frac{1}{4\alpha^2}.
\eear
\setcounter{equation}{0}
\section{Four-dimensional Taub-NUT and Taub-NUT-AdS$_4$ Thermodynamics with GUP}
In this section, we will show
how the extended forms of HUP
modify the Hawking temperature
of the four-dimensional
Taub-NUT and the
Taub-NUT-AdS$_4$ black hole.
\subsection{Hawking Temperature with GUP in four-dimensional Taub-NUT space}
\label{TaubNutGUP} Here we
apply GUP~(\ref{GUP}) to
Taub-NUT space and check how it
modifies the Hawking
Temperature from HUP.
Using~(\ref{ineqGUPp}) with
$\gamma=1$, we can infer that
Hawking Temperature is given by
\begin{eqnarray}\label{GUPTem}
T_{\rm GUP}=\left(\frac{1}{4\pi}\right)\frac{r_{+}}{2\alpha^2
\l_{p}^2} \Biggr[ 1 - \sqrt{1-\frac{4\alpha^2 \l_{p}^2}{r_{+}^2}}
\Biggr] \hbar c\,,
\end{eqnarray}
where we replaced $\Delta x$ by
$r_+$,  chose the left side
inequality for the minimum
energy. The absolute
minimum value in the position
uncertainty (\ref{GUPxmin})
satisfies
\begin{eqnarray}\label{GUPxxmin}
r_{+} \ge 2 \alpha l_{p}\,.
\end{eqnarray}
When $r_+ \gg \alpha l_p$ (the
large black hole), the Hawking temperature is given
by~\cite{Park:2007az,Hawking:1974sw,Myers:1986un}
\begin{eqnarray}\label{GUPTemApro}
T_{\rm GUP} \simeq \frac{1}{4\pi} \Biggr[\frac{1}{r_{+}}
 + \frac{\alpha^2 \l_{p}^2}{r_{+}^3} \Biggr]\hbar c \,.
\end{eqnarray}

Since the second term in
(\ref{GUPTemApro}) gives always
a positive correction, the
magnitude of Hawking
temperature with GUP is always
bigger than that of Hawking
temperature from HUP. Therefore,
the temperature of the Taub-NUT
black hole due to the GUP
increases like that of
Schwarzschild black hole so
that the Hawking temperature is faster by virtue of the generalized uncertainty
principle up
to a minimum radius
$r_{min}=2\alpha l_p$.
This lower bound $r_{min}$
comes from the second term in
the GUP (\ref{GUP}). Thus, the
GUP can prevent the Taub-NUT
black hole from complete
evaporation which arises from
emitting black body radiation
at the Hawking temperature,
i.e. the Taub-NUT black hole
evaporation stops at $r_{min}$.
In fact, using rather generic
and model-independent
considerations, the GUP with
the similar result of string
theory is
found~\cite{Veneziano:1986zf,Gross:1987ar,Amati:1988tn,Konishi:1989wk}.
The scale of minimum radius
$r_{min}$ itself is not
explicitly determined by the
GUP. With the help of string
theory, such a scale is
decided, i.e. this scale can be
considered as a melting scale
of the Taub-NUT black hole
which is followed by the string
phase~\cite{Susskind:1993ws,Park:2007az}.

The GUP (\ref{GUP}) gives an
interpolation between the
quantum mechanical limit and
the quantum gravity
limit~\cite{Bolen:2004sq}. When
$\Delta p \ll \hbar /(\alpha
l_p)$, we get the quantum
mechanical limit. The quantum
gravity limit is calculated
when $\Delta p \simeq \hbar
/(\alpha l_p)$.

\subsection{Hawking Temperature with the dual form of the GUP in the Taub-NUT-AdS$_4$ space}
\label{EUPAds4} Since the GUP
does not have any upper bound
on the maximum uncertainty in
the position, the GUP can not
be naively applied to the (A)dS
limit~\cite{Amati:1988tn,Konishi:1989wk,Myung:2006qr,Park:2007az}.
In this subsection, we will
examine the corrections to the
Hawking Temperature due to (\ref{EUP}) in the Taub-NUT-AdS$_4$
space. The symmetric generalized uncertainty principle will be
discussed in following
subsection.

The uncertainty in the energy
of the emitted particle is
given by
\begin{eqnarray}\label{TNAEUP}
\Delta E \simeq c \Delta p \simeq \biggr[ \frac{\gamma}{\Delta x} +
\frac{\beta^2}{l_{p}^2}\Delta x \biggr] \hbar c\,.
\end{eqnarray}
When we identify the parameter $\beta$ with the second term in (\ref{TNAT})
as
\begin{eqnarray}\label{beta}
\beta^2 \equiv \frac{3\,l_{p}^2}{l^2}\,,
\end{eqnarray}
we reproduce the Hawking
temperature of the
Taub-Nut-AdS$_4$ black hole
(\ref{TNAT}) by (\ref{EUP}).

(\ref{EUP}) also gives
an interpolation between the
two different quantum
mechanical and the quantum
gravity
limits~\cite{Bolen:2004sq}. The
quantum mechanical limit is
computed when $\Delta x \ll
l_p/\beta$, which leads to
$\beta\ll 1$. When $\Delta
x\simeq l_p/\beta$, the quantum
gravity limit is achieved and
the quantum gravitational
effects give the manifest
contribution at very large
distance. Thus, (\ref{EUP}) has an
interpolation between two
thermal limits.

As mentioned above, in contrast
to four-dimensional Taub-NUT
black hole, the
Taub-NUT-AdS$_4$ solution has
the two thermodynamical limits.
In fact, they arise from the
two limiting relations between
the position and the momentum
of (\ref{EUP}), i.e. $\Delta
p\simeq \hbar /\Delta x$ in
quantum mechanical (low energy)
limit and $\Delta p\simeq \hbar
\Delta x/l_p^2$ in quantum
gravity (high energy) limit.
Furthermore, the Hawking
temperature of the
Taub-NUT-AdS$_4$ is obtained by
the high energy limit of the
(\ref{EUP}), which seems to reflect a
quantum gravitational nature.
However, even if the HUP
suffices to derive the
four-dimensional Taub-NUT black
hole thermodynamics, there does
not exist consistent
formulation of string theory in
even Schwarzschild geometry as
well as in the Taub-NUT space.
Therefore, the Hawking
temperature of the
Taub-NUT-AdS$_4$ solution seems
to have a different derivation
from that of the
four-dimensional Taub-NUT
metric. One may apprehend why
only the Taub-Nut-(A)dS black
hole may allow AdS/CFT
correspondence in spite of
their distinctive properties
such as ALF and breakdown of
the area theorem of the
horizon.
\subsection{Hawking temperature with symmetric generalized uncertainty principle in Taub-NUT-AdS$_4$ space}
Substituting the parameter $\beta$ (\ref{beta}) in the inequality (\ref{ineqp}),
we find the Hawking temperature
\begin{eqnarray}\label{GEUPTAds}
\tilde{T}_{{\rm AdS}} = \left(\frac{1}{4\pi}\right) \frac{
r_+}{2\alpha^2 l_{p}^{2}} \Biggr[ 1 - \sqrt{1-\frac{4 \alpha^2
l_{p}^{2}}{r_{+}^{2}}\biggr[\gamma + \frac{3 r_{+}^2}{l^2}\biggr] }
~\Biggr] \hbar c\,.
\end{eqnarray}
Since we require the square root in (\ref{GEUPTAds}) to be real,
we get the minimum radius $r_+$
\begin{eqnarray}\label{GEUPrmin}
r_+ \ge \frac{2 \alpha l_{p}\sqrt{\gamma}}{\sqrt{1-{12 \alpha^2 l_{p}^{2}}/{l^2}}}\,.
\end{eqnarray}
When one keeps the first
leading power in $1/r_+$, i.e.
semi-classical regime $\alpha
l_p \ll r_+ \ll l$, the Hawking
temperature with
(\ref{GEUPTAds}) goes back to
result leading to (\ref{EUP}). When one keeps
the second leading power in
$1/r_+$, the temperature
(\ref{GEUPTAds}) is written as
\begin{eqnarray}\label{4GEUPTAdsEUP}
\tilde{T}_{\rm AdS} \simeq \frac{1}{4\pi}
\Biggr[ \Biggr(\gamma+\frac{6\gamma \alpha^2 l_p^2}{l^2}\Biggr)\frac{1}{r_{+}}
+ \frac{3r_{+}}{l^2}
+\frac{\gamma^2\alpha^2 l_p^2}{r_{+}^3}~\Biggr] \hbar c\,.
\end{eqnarray}
Here, all terms in the bracket
are positive, and the Hawking
temperature of the
Taub-NUT-AdS$_4$ with (\ref{GEUPTAds})
increases. Then,
the black hole decays up to the
minimum radius
$r_{min}$~(\ref{GEUPrmin}).
Also the Hawking temperature
grows up according to
increasing of the NUT charge.

\setcounter{equation}{0}
\section{Taub-NUT and Taub-NUT-(A)dS Thermodynamics in Arbitrary Even
Dimensions} In this section, we
will first show that the
Hawking temperature with the
extended versions of HUP can be
obtained in arbitrary even
dimensional Taub-NUT-AdS space.
Next, we will take the limit of
cosmological constant going to
zero and get the temperature of
the Taub-NUT black hole.
Finally, we will obtain the
Hawking temperature of the
Taub-NUT-dS space through an
analytic continuation of the
cosmological constant.
\subsection{Taub-NUT-AdS Thermodynamics with symmetric generalized uncertainty principle}
The Taub-NUT-AdS metric in
higher dimensions has the
following
form~\cite{Awad:2000gg,Clarkson:2002uj}
\begin{eqnarray}\label{dTNAmetric}
ds^{2}=-f(r)\biggr( dt + 2n \sum_{i=1}^{k}\cos(\theta_i) \theta d\phi_i \biggr)^{2}
       +\frac{dr^{2}}{f(r)} + (r^{2}+n^{2})d\Omega_{d-2} \,,
\end{eqnarray}
where $(d+1) = 2k+2$ is the total number of dimension and
$d\Omega_{d-2}$ is the area of the unit $S^{d-2}$. The metric
function $f(r)$ has the general form
\begin{eqnarray}\label{dTNAfr}
f(r)=\frac{r}{(r^{2}+n^{2})^2}
\int^{r}\Biggr[\frac{(s^2+n^2)^k}{s^2}+\frac{2k+1}{l^2}\frac{(s^2+n^2)^{k+1}}{s^2}\Biggr]ds
-\frac{2mr}{(s^2+n^2)^k}\,.
\end{eqnarray}
The Hawking temperature can be written using (\ref{ghtemp}) as
\begin{eqnarray}\label{dTNAT}
T_{\rm H(AdS)} = \left(\frac{d-2}{4\pi}\right) \biggr[
\frac{1}{r_{+}} +
\Biggr(\frac{d}{d-2}\Biggr)\frac{n^2+r_{+}^2}{l^2\,r_{+}}
\biggr]\hbar c\,.
\end{eqnarray}
Using the relation
(\ref{EUP}) and following the
same method as in subsection
\ref{EUPAds4}, we find that we
can reproduce the above Hawking
temperature (\ref{dTNAT}) from
(\ref{EUP}) by identifying
\begin{eqnarray}\label{dbeta}
\gamma_{d} \equiv
1+(\frac{d}{d-2})\frac{n^2}{l^2}~,
~~~\beta^2 \equiv
\Biggr(\frac{d}{d-2}\Biggr)\frac{\,l_{p}^2}{l^2}
\,.
\end{eqnarray}

 Similarly the Hawking
temperature coming from
(\ref{ineqp}) in the
Taub-NUT-AdS space is given by
\begin{eqnarray}\label{dGEUPTAds}
\tilde{T}_{\rm AdS} = \left(\frac{d-2}{4\pi}\right) \frac{
r_+}{2\alpha^2 l_{p}^{2}} \Biggr[ 1 - \sqrt{1-\frac{4 \alpha^2
l_{p}^{2}}{r_{+}^{2}}\biggr[\gamma_{d} +
\Biggr(\frac{d}{d-2}\Biggr)\frac{r_{+}^2}{l^2}\biggr] }
~\Biggr] \hbar c\,,
\end{eqnarray}
and the absolute minimum $r_+$
\begin{eqnarray}\label{dGEUPAdsrmin}
r_+ \ge
\frac{2 \alpha l_{p} \sqrt{\gamma_{d}}}{\sqrt{1-(\frac{d}{d-2})\frac{4 \alpha^2 l_{p}^{2}}{l^2}}} \,.
\end{eqnarray}
Keeping the second leading power in $1/r_+$, the temperature with (\ref{dGEUPTAds})
is found as
\begin{eqnarray}\label{dGEUPTAdsEUP}
\tilde{T}_{\rm AdS} \simeq \left(\frac{d-2}{4\pi}\right) \Biggr[
\Biggr\{\gamma_{d}+\Biggr(\frac{d}{d-2}\Biggr)\frac{2 \alpha^2 l_p^2}{l^2}
\Biggr\}\frac{1}{r_{+}} +
\Biggr(\frac{d}{d-2}\Biggr)\frac{r_{+}}{l^2}+\frac{\gamma_{d}^2\alpha^2
l_p^2}{r_{+}^3}~\Biggr] \hbar c\,.
\nonumber\\
\end{eqnarray}
This shows that the corrections due to (\ref{ineqp}) increase the Hawking
temperature without depending on the dimensions of the Taub-NUT-AdS
since all corrections in (\ref{dGEUPTAdsEUP}) are positive.

Now, taking the limit of cosmological constant approaching zero, we
will obtain the Hawking temperature with the GUP in the arbitrary
dimensional Taub-NUT space. When $l\rightarrow\infty; \gamma_{d}\rightarrow 1$, the
temperature~(\ref{dGEUPTAds}) reaches to the Hawking temperature
with the GUP
\begin{eqnarray}\label{dTaubnut}
T_{\rm GUP} = \left(\frac{d-2}{4\pi}\right) \frac{ r_+}{2\alpha^2
l_{p}^{2}} \Biggr[ 1 - \sqrt{1-\frac{4 \alpha^2
l_{p}^{2}}{r_{+}^{2}}} ~\Biggr] \hbar c\,,
\end{eqnarray}
and the absolute minimum $r_+$
has the same form
as~(\ref{GUPxxmin})
independently on the
dimensions. Then, for the large
black hole, the
temperature~(\ref{dTaubnut})
approaches
\begin{eqnarray}\label{GUPdTemApro}
T_{\rm GUP} \simeq \left(\frac{d-2}{4\pi}\right)
\Biggr[\frac{1}{r_{+}}
 + \frac{\alpha^2 \l_{p}^2}{r_{+}^3} \Biggr]\hbar c \,.
\end{eqnarray}
Thus, one finds that the Hawking temperature with the GUP in the
arbitrary dimensional Taub-NUT space also increases.

\subsection{Taub-NUT-dS Thermodynamics with symmetric generalized uncertainty principle}
Employing analytic continuation
of the cosmological parameter
$l \rightarrow il$, and
substituting in
~(\ref{dTNAT})--(\ref{dGEUPTAdsEUP})
in the AdS space, one can
obtain several results in the
dS space as follows: the
Hawking temperature is given by
\begin{eqnarray}\label{dTNdsT}
T_{\rm H(dS)} = \left(\frac{d-2}{4\pi}\right) \biggr[
\frac{\lambda_{d}}{r_{+}} \,-
\Biggr(\frac{d}{d-2}\Biggr)\frac{r_{+}}{l^2}
\biggr]\hbar c\,,
\end{eqnarray}
where $\lambda_{d} \equiv
1-(\frac{d}{d-2})\frac{n^2}{l^2}$,
the $\beta$ and the Hawking
temperature with (\ref{ineqp}) becomes
\begin{eqnarray}\label{dbeta}
\beta^2 \equiv
-\Biggr(\frac{d}{d-2}\Biggr)\frac{\,l_{p}^2}{l^2}\,,
\end{eqnarray}
and
\begin{eqnarray}\label{dGEUPTds}
\tilde{T}_{\rm dS} = \left(\frac{d-2}{4\pi}\right) \frac{
r_+}{2\alpha^2 l_{p}^{2}} \Biggr[ 1 - \sqrt{1-\frac{4 \alpha^2
l_{p}^{2}}{r_{+}^{2}} \biggr[\lambda_{d} -
\Biggr(\frac{d}{d-2}\Biggr)\frac{r_{+}^2}{l^2}\biggr] }
~\Biggr] \hbar c\,.
\end{eqnarray}
The absolute minimum $r_+$ is given by
\begin{eqnarray}\label{dGEUPdsrmin}
r_+ \ge \frac{2 \alpha l_{p} \sqrt{\lambda_{d}}}{\sqrt{1+(\frac{d}{d-2})\frac{4
\alpha^2 l_{p}^{2}}{l^2}}} \,,
\end{eqnarray}
and  for large $r_+$, we have
\begin{eqnarray}\label{dGEUPTdsEUP}
\tilde{T}_{\rm dS} \simeq \left(\frac{d-2}{4\pi}\right) \Biggr[
\Biggr\{\lambda_{d}-\Biggr(\frac{d}{d-2}\Biggr)\frac{2 \alpha^2 l_p^2}{l^2}
\Biggr\}\frac{1}{r_{+}}\, -
\Biggr(\frac{d}{d-2}\Biggr)\frac{r_{+}}{l^2}
+\frac{\lambda_{d}^2\alpha^2 l_p^2}{r_{+}^3}~\Biggr] \hbar c\,.\nonumber\\
\end{eqnarray}

The magnitude of the Hawking
temperature with (\ref{ineqp}) is always
bigger than that of Hawking
temperature with (\ref{EUP}) like the
case of AdS space.
Therefore, black hole with symmetric generalized uncertainty principle
temperature evaporates faster
up to a minimum radius
$r_{min}$. Since
the minimum radius
$r_{min}$~(\ref{dGEUPdsrmin})
must be real, the maximum value
for a NUT charge is given by
\begin{eqnarray}\label{dsnmax}
n^2 \leq \, \biggr(\frac{d-2}{d}\biggr)l^2 \equiv n_{max}^2,
\end{eqnarray}
and the maximum value for $r_{+}$ is
\begin{eqnarray}\label{dsrmax}
r_+ \leq \, \sqrt{\biggr(\frac{d-2}{d}\biggr)l^2-n^2} \equiv r_{max} \,.
\end{eqnarray}
In contrast to the Taub-NUT-AdS, the Hawking temperature of the
Taub-NUT-dS decreases according to increasing of the NUT charge up to
the upper bound $n_{max}$ ~(\ref{dsnmax}).

\section{Conclusion}
Considering the Taub-NUT-(A)dS spaces and employing the extended
versions of the HUP based on the string theory, we showed that these
versions give rise to their thermal quantities. We also found that
two thermodynamical limits of the Taub-NUT-(A)dS metric are the low
energy limit and the high energy limit like S-(A)dS-like black hole.
This seems to reflect that the thermodynamic origin of the
Taub-NUT-(A)dS is different from that of the Taub-NUT metric. We
also obtained the result that the generalized forms of the HUP
increase the Hawking temperature of Taub-NUT-(A)ds black as in the
case of the S-(A)dS like black hole.

We added one more instance of the Hawking temperature with the
property that the flat case can be approached with the ordinary
Heisenberg uncertainty principle, whereas the (A)dS space must
utilize the extended forms of the Heisenberg uncertainty principle.
Of course, deriving the thermodynamic quantities from the
generalized Heisenberg uncertainty  principle and having two limits
does not necessarily imply holography, and definite conclusion
awaits a full quantum states of (A)dS space.  Nevertheless, the
extended uncertainty principle is directly connected with the finite
infrared (IR) boundary of the (A)dS space, and from this, we
speculate that Taub-NUT-(A)dS might admit holographic dual
description as in the case of Schwarzschild-(A)dS black hole.

\section*{Acknowledgements}

This work was supported by the Korea Research Foundation Grant funded
by the Korean Government (KRF-2008-357-C00018), the BK 21 project of
the Ministry of Education and Human Resources Development, Korea, and
by the Science Research Center Program of the Korea Science and
Engineering Foundation through the Center for Quantum
Spacetime(CQUeST) of Sogang University with grant No. R11-2005-021.


\end{document}